Shaking a container full of perfect liquid; a tractable case, a torus shell, exhibits a virtual wall

J.H. Hannay,
H.H. Wills Physics Laboratory,
Tyndall Avenue,
Bristol BS8 1TL, UK.

*Abstract*
Manipulation ('shaking') of a rigid container filled with incompressible liquid starting from stationary generally results in some displacement, or mixing, of the liquid within it.  If the liquid also has zero viscosity, a 'perfect' or 'Euler' liquid, Kelvin's theorems dramatically simplify the flow analysis.  Response is instantaneous; stop the container and all liquid motion stops.  In fact an arbitrary manipulation can be considered as alternating infinitesimal translations and rotations of the container.  Relative to the container, the liquid is stationary during every translation.  Infinitesimal rotations (an infinitesimal vector along the rotation axis) resolve into three orthogonal components in the container frame.  Each generates its own infinitesimal liquid displacement vector field.  However these are rarely tractable, and their combined consequences are obscure.  Rather than a volume flow, a surface flow in 3D is considerably easier, the liquid slipping freely in a shell, sandwiched between two nested closed surfaces with constant infinitesimal gap.  The closedness avoids extra boundaries.  The two dimensionality admits a scalar streamfunction determined by the container angular velocity vector.  Manipulation of the container angular velocity at will, leads (except for a sphere) to an infinitely rich variety of area preserving re-configurations of the liquid.  Even for a sphere, any chosen point of the liquid can be moved, in the shell container frame, to any other point, and one would expect the same in general.  However, for a torus with a small enough hole (diameter<0.195 torus diameter), there exists a virtual wall, a hypothetical axial cylinder intersecting the torus.  No matter how the torus is manipulated, liquid inside the cylinder stays inside; outside stays outside.  The analysis, solving for the streamfunction, is based on the relative vorticity, and the conformal mapping of a torus to a (periodic) rectangle, which lead to a fairly simple convolution integral formula for the flow.

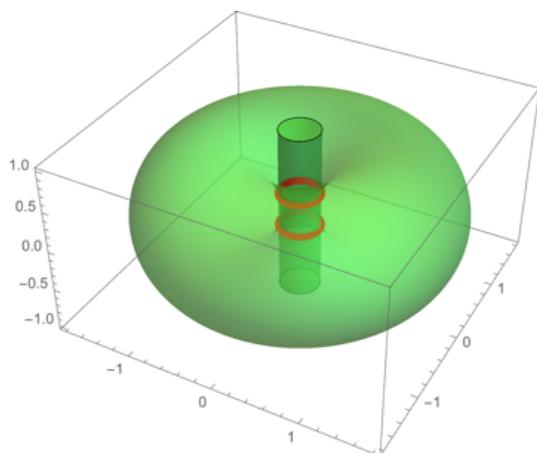

Fig 1.  There is a virtual cylinder construction associated with surface flow on a rigid torus (driven by torus manipulation) provided the torus is fat enough. The cylinder is somewhat

wider than the hole in the torus and therefore intersects it in two parallel circles or 'twin rings' as shown. This hypothetical cylinder constitutes a virtual wall: no manipulation whatever of the torus can cause any liquid to cross it. Liquid starting outside stays outside, inside stays inside.

*Motivation and results, pictorially*

Manipulation ('shaking') of a rigid container filled with incompressible liquid starting from stationary generally results in some displacement, or mixing, of the liquid within it. This applies even if the liquid has zero viscosity – a perfect liquid, treated here. One might question this statement - think of a liquid filled spherical container – a simple rotation of the container about its centre obviously does nothing at all to the liquid which remains stationary, slipping frictionlessly. But, in a frame of reference fixed in the container, the liquid has moved – by a rigid counter rotation. This same tendency is present for a general container shape. Kelvin's theorems (recalled at the end of this section) dramatically reduce the flow analysis to geometry (rather than dynamics). From his Kinetic energy theorem, response is instantaneous; stop the container and all liquid motion stops. In fact an arbitrary manipulation can be considered as alternating infinitesimal translations and rotations of the container. Relative to the container, the liquid is stationary during every translation. For infinitesimal rotations, however (except around a symmetry axis), the infinitesimal liquid displacement fields are much less trivial. There is a separate such field for each of three rotation axis directions in the container frame and the consequences of a sequence of them are obscure.

Striving for yet more simplicity, reduced dimensionality is often attractive; here realized by a rigid container that 'sandwiches' a liquid in a shell of uniform infinitesimal thickness so that the flow is confined to a surface. The analytical benefit of 2D is the existence of a scalar streamfunction and Kelvin's Constant circulation theorem relates this function in the next section, to the container angular velocity. A boundaryless 'closed' surface is simplest. Except for a sphere (spherical shell) for which any sequence of rotations leads to a 'rigid' relative displacement of the liquid, the freedom to manipulate the container angular velocity at will, leads to an infinitely rich variety of area preserving re-configurations of the liquid. It is natural to expect that at least, in common with the sphere, any chosen point of the liquid can be moved to any chosen point in the shell container. This will prove not to be the case (fig 1), a fat torus providing a striking counterexample. Less symmetric shell shapes than the torus, however, seem very likely to admit, like the sphere, the arbitrary point to arbitrary point displacement that is lacking in the fat torus.

Specializing, from now on, to the torus, natural axes in the torus frame of reference are the symmetry axis **a**, and two orthogonal 'diametral' ones. The relative flow consequence of turning the torus about its symmetry axis is obvious. The liquid merely counter rotates rigidly with frictionless slipping. Rotation about a diametral axis at angular velocity $\omega_\perp$ causes a more elaborate flow pattern that needs calculating in the remaining sections. (The streamline pattern is independent of $|\omega_\perp|$). Some features are forced by symmetry and have an important consequence later. The two circles in the mirror symmetry plane normal to the turning axis (figs 2 and 4) are necessarily streamlines, though with uneven flow speed.

There is thus no flow across either of them, and, by incompressibility, there can be no net flux across any longitude circle $\phi$=const (fig 2). The same zero net flux also applies to every latitude circle $\theta$=const. Again, zero flux for a single circle suffices, say the largest $\theta=\pi$, which has zero net flux by the $\pi$ rotational symmetry of the flow about the diametral turning axis.

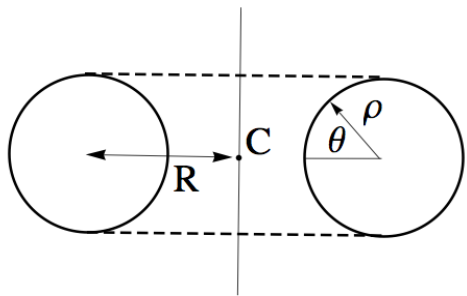

Fig 2. Vertical cross section bisecting a horizontal torus, defining its shape and size. The ratio $\rho/R$ describes the torus shape, from thinnest to fattest: $0<\rho/R<1$. The ratio can be neatly expressed in solid angle terms: it is the proportion of all $4\pi$ solid angle taken by the torus when viewed from its centroid C. Natural coordinates on the torus are: a latitude-like one $\theta$ (but with $-\pi<\theta<\pi$) as shown, and a longitude-like one $\phi$ (with $-\pi<\phi<\pi$) not shown, measuring azimuth angle around the symmetry axis.

In case a qualitative description of the outcome suffices, or perhaps motivates the proper calculation, it can be supplied now, pictorially. There is a difference in the topology of the diametral turning flow for a thin torus, $\rho/R<0.674...$ and a fat torus, $\rho/R>0.674$. The flow streamlines on the torus have the respective topologies of contours of $\omega_\perp\cdot\mathbf{r}$ and of $\omega_\perp\cdot\mathbf{n}$, where $\mathbf{n}$ is the unit surface normal vector at $\mathbf{r}$. These separate types of contours are shown in fig(3).

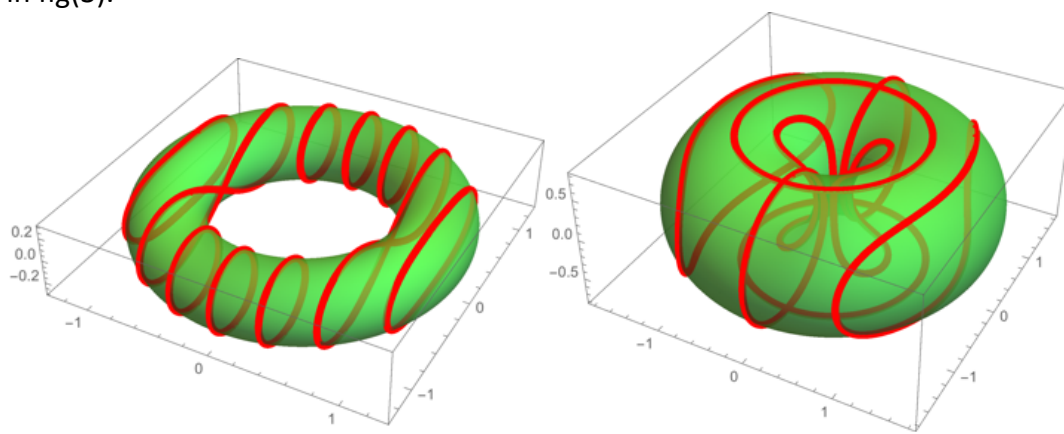

Fig 3. Turning a torus about a diametral axis (horizontal, at any angular velocity $\omega_\perp$) produces flow of a perfect liquid confined to its surface. For visual clarity and analytical assimilation, the lines shown here correctly capture the flow topology but have the oversimplified analytical forms stated. Evidently the two pictures differ. Left: Thin torus

($\rho/R=0.3<0.674..$) with contours of $\omega_\perp \cdot \mathbf{r}$ drawn. They are somewhat unrealistic geometrically since the true streamline loops are not typically planar. Right: Fat torus ($\rho/R=0.8>0.674..$) with some contours of $\omega_\perp \cdot \mathbf{n}$ drawn, where $\mathbf{n}$ is the unit surface normal vector. The important two horizontal circle streamlines, the 'twin rings', are however, unrealistically located at the top and bottom – they should be closer together as in fig 1, the intersections of the torus with a fairly narrow axial cylinder. The cylinder radius for thinner tori approaches the hole radius, the twin rings coalescing and annihilating for $\rho/R=0.674$. The twin rings are shown correctly located in fig 4.

A fat torus has twin horizontal circular streamlines, 'twin rings', intersections of the torus with an axial cylinder as in fig 1. Crucially, and obviously by axial symmetry, this same twin ring pair, applies for any diametral axis of turning of the torus. And for axial turning of the torus, the twin pair is still also obviously present, included among the entirety of horizontal circular streamlines. That means: no liquid can be made to cross the twin pair under any manipulation whatever of the torus; liquid inside the intersecting cylinder stays inside, outside stays outside. Fig 4 shows the computed wall circles for several tori of different ratios $\rho/R$.

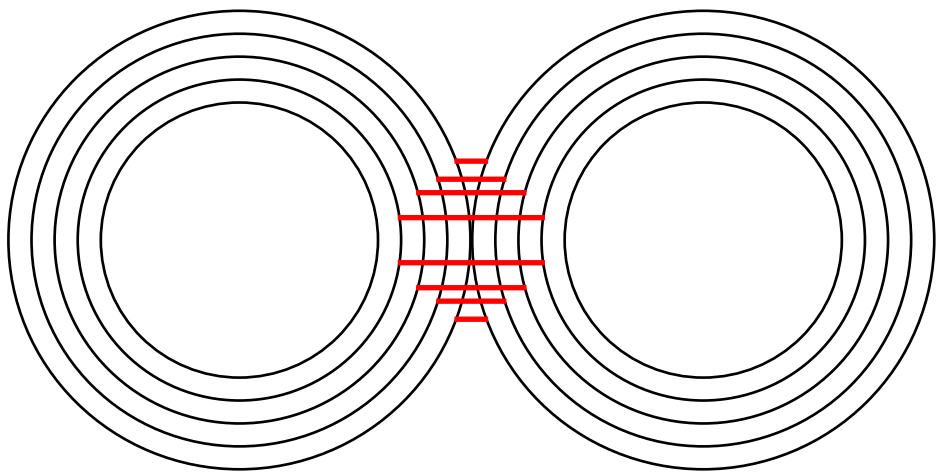

Fig 4. Five torus pictures in one. Each shows the bisecting cross section of a torus in black; the ratios $\rho/R$ being 0.6, 0.7, 0.8, 0.9, 1.0 ($\rho$ is the black circle radius, and $2R$ the centre-centre distance). Also shown for each torus, in red, is the edge-on projection (not cross-section) of the twin horizontal virtual wall rings lying in the torus surface. The exception is the thinnest torus shown with ratio $\rho/R=0.6<0.674$, which has no virtual wall rings. The fattest possible torus $\rho/R=1$ (closed hole) has twin rings with vertical separation $2\rho \sin(0.352)$, computed from (12). The separation is graphed in fig 5.

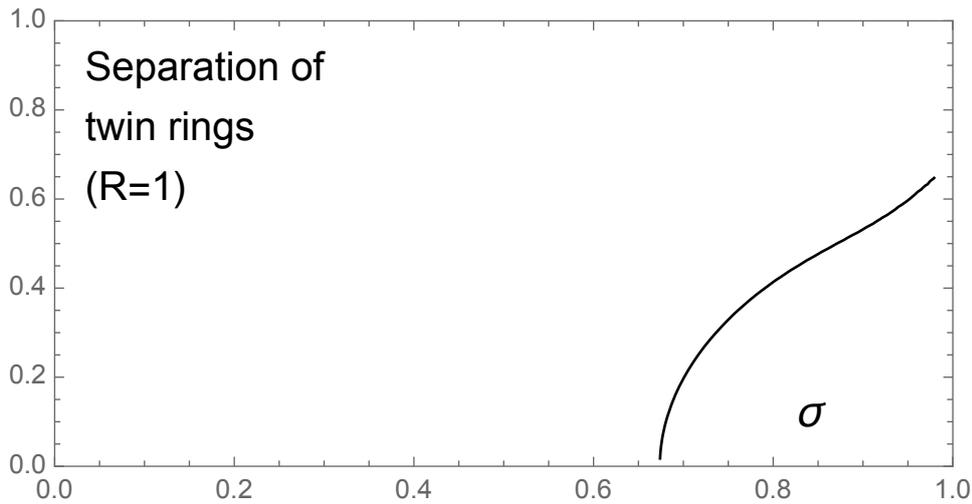

Fig 5.  Graph showing how the separation of the twin virtual wall rings in fig 4 depends on the fatness $\sigma=\rho/R$ of the torus (taking $R$=1).

Concluding this section, Kelvin's two theorems [Lamb 1932] for a perfect liquid (incompressible and having zero viscosity) should be specified.  They apply when boundary motion alone drives the liquid motion, so this covers the 'shaking' of a full container of perfect liquid.  If started from stationary his Constant convected circulation theorem states that the circulation (the line integral of velocity around a hypothetical test circuit within the liquid) remains zero for any test circuit.  This will relate the (Laplacian of the) streamfunction, in the next section, to the local normal component of container (shell) angular velocity.  His Kinetic energy theorem dictates that the instantaneous motion of the liquid is that with the least total kinetic energy consistent with the prescribed normal component of motion of the boundary.  This proves the instantaneous response mentioned earlier; there is no history dependence.

*Relative flow streamfunction*

Since the incompressible liquid shell has uniform infinitesimal thickness the relative surface flow is divless. Such a flow has the property that if, at every point, the relative velocity tangent vector is hypothetically twisted by a right angle about the local surface normal vector (all in the same sense), that new vector field is curlless.  It is therefore the gradient of a scalar function *s* (possibly globally multivalued but always locally meaningful – more correctly the gradient is a closed one–form).  The div of this right angle-twisted vector field $\nabla.(\nabla s)\equiv\nabla^2 s$ equals the curl of the relative velocity field which lies along the surface normal direction, that is, its relative vorticity.  The important points are that this relative vorticity is not zero (the flow is not irrotational in a rotating frame), but that it is explicitly known.  It is simply expressed in terms of the angular velocity vector $\omega$ of the rigid shell and the local normal unit vector **n** of the shell.  At all times and at all positions

$\nabla^2 s=-2\omega.\mathbf{n}$ (1)

The justification is as follows. In general for a perfect liquid the vorticity, the curl of liquid velocity in space (as opposed to relative vorticity), is well known to remain permanently zero everywhere if it starts so. But its interpretation is subtle for a flow confined to a moving surface, and it is safest to use the equivalent integral version, Kelvin's Constant convected circulation theorem: any chosen closed circuit carried, or convected, by the liquid motion in 3D, has unchanging line integral of velocity around it, in this case zero, since the liquid starts stationary. In particular, infinitesimal circuits have permanently zero circulation. This circulation is twice the infinitesimal circuit area times the dot product of **n** with the local angular velocity vector of the liquid in space. But this angular velocity in space equals the angular velocity ω of the shell plus the angular velocity of the liquid relative to the shell frame of reference, $\lambda$**n**, say, since it must point along the local surface normal. So one has 2(ω+$\lambda$**n**).**n**=0, yielding $\lambda$=−2(ω.**n**) as the relative vorticity of the liquid on the right hand side of (1).

As indicated before, since the flows produced by components of the torus angular velocity simply superpose, it is convenient to treat them separately, resolving ω into axial and diametral vectors: ω = $ω_a$ + $ω_⊥$ where, in terms of the unit axial vector **a**, $ω_a$=**a**(ω.**a**) and $ω_⊥$=ω−**a**(ω.**a**). The streamfunction corresponding to $ω_a$ is obvious without reference to the equation (1). The rigid counter rotation means $\rho \partial s/\partial \theta = -(\omega.\mathbf{a})(R - \rho\cos\theta)$. The streamfunction corresponding to $ω_⊥$ is to derived in the next sections. It requires the relative vorticity expressed in coordinates

$$-2(\omega_⊥ .\mathbf{n}) = 2\ \omega_⊥ \cos\theta \cos\phi \qquad (2)$$

with the point $\theta = \phi = 0$ defined as lying on the 'hole waist' circle of radius $R - \rho$ at the azimuth, say, where $ω_⊥$ **and n** are antiparallel. From here on, the torus frame of reference will be understood, so the qualification 'relative' will be dropped.

This paragraph is a pre-emptive dismissal of a potential complication. The velocity field that derives from a given vorticity field is generally not unique. On a torus there is a two parameter family of undetermined div-less, curl-less flows, briefly described now, but dismissed as zero by the symmetry of the flow from the diametral turning of the torus. One div-less, curl-less, flow has equal circulation around all latitude circles (so speed inversely proportional to circle radius). At position vector **r**, the velocity is **v(r)** ∝ (**r** ∧ **a**)/ (**r** ∧ **a**)$^2$. Another such flow is along longitude circles. It is obtained by rotating each previous vector by a right angle (in the same sense) about the local normal vector **n**. These two flows [Klein 1932], and therefore also any linear combination of them, exhaust all the div-free, curl-free flows allowed by torus topology. The two div-less, curl-less flows described have non zero flux respectively across longitude circles, and across latitude circles. But as noted in the previous section diametral turning of the torus causes zero net flux across all longitude and latitude circles. The desired flow, constructed from a vortex lattice below, has zero fluxes built in.

*Conformal map of a torus*

The task is to invert the Laplacian to obtain *s* in terms of the vorticity $-2(\boldsymbol{\omega}_\perp \cdot \mathbf{n})$. The difficulty is that the torus is not flat, but it can be mapped in a standard way to a flat rectangle with periodic boundary conditions on which the Laplace inversion process is more familiar. Once found, the stream function *s* can be carried back directly onto the torus by the inverse mapping. The latitude and longitude circles on the torus (fig 2) map to horizontal and vertical straight lines on the rectangle, but simply taking $(\phi,\theta)$ as cartesian coordinates forming the rectangle is not adequate.

Only for a conformal mapping (angle preserving, thus locally shape preserving) is the Laplacian of a function proportional (via the local area ratio) to the Laplacian of the mapped function. For this local shape preservation, the rectangle coordinates need to be $(\phi,\theta')$ where $\theta'(\theta)$ is a specific function (4) below, a result apparently [1] going back to Kirchhoff. Since all the coordinate meshes on the rectangle have the same shape, the function must be such that latitude lines drawn on the torus at equal intervals of values of $\theta'$ (rather than $\theta$) form a mesh pattern on the torus in which all the mesh shapes are the same (with varying size), fig(6). The torus latitude lines have differing circumferences proportional to $(1-\sigma\cos\theta)$ where $\sigma=\rho/R$. For mesh shape matching one therefore needs

$$d\theta/d\theta' = (1 - \sigma \cos\theta)/\sqrt{1-\sigma^2} \qquad (3)$$

where the denominator ensures that, when integrated, $\theta'$ runs from $-\pi$ to $\pi$ like $\theta$. The common aspect ratio of the mesh shapes $d\phi \times d\theta'$ can be read off from the case $\theta = \pi/2$ where $d\theta/d\theta'=1/\sqrt{1-\sigma^2}$, so $d\phi \times d\theta' = d\phi \times d\theta/\sqrt{1-\sigma^2}$ which on the torus has aspect ratio $\rho/R/\sqrt{1-\sigma^2} = \sigma/\sqrt{1-\sigma^2}$. This is the aspect ratio of the rectangle conformally mapped from the torus with $\sigma=\rho/R$. The rectangle is a square at $\sigma = 1/\sqrt{2}$; it is taller than wide for fatter tori, or wider than tall for thinner tori. The overall size of the rectangle is arbitrary; for definiteness, that with horizontal width $2\pi R$ is chosen, the vertical height then being $2\pi R\sigma/\sqrt{1-\sigma^2}$.

The solution of the differential equation (3) is

$$\theta' = 2\tan^{-1}\left(\sqrt{\frac{1+\sigma}{1-\sigma}}\tan\tfrac{1}{2}\theta\right) \quad \text{or in reverse:} \quad \theta = 2\tan^{-1}\left(\sqrt{\frac{1-\sigma}{1+\sigma}}\tan\tfrac{1}{2}\theta'\right) \qquad (4)$$

(This formula for conformal torus mapping is also well known in other physical contexts, relativistic aberration of light, for example).

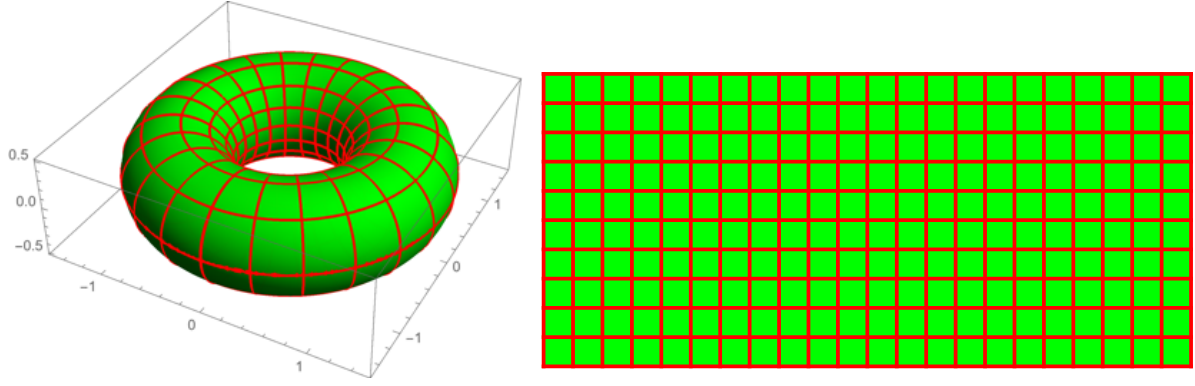

Fig 6. Left: the torus with the horizontal, latitude, circles drawn closer together on the inside proportionally to the closeness of vertical, longitude, circles. This creates meshes all having the same shape, though not size. Right: those same horizontal and vertical circles mapped to straight lines with that same mesh shape, suitably resized. This defines the conformal (local shape preserving) map (4) of the torus (here with $\sigma=\rho/R$ =0.5) onto a periodic rectangle. It has a definite aspect ratio $\sigma/\sqrt{1-\sigma^2}$.

To write down the vorticity on the rectangle one needs the ratio of areas of mesh elements $d\phi\, d\theta'$ on the torus and the rectangle, namely $R d\phi \rho d\theta(1-\sigma\cos\theta)$ to $R^2 d\phi\, d\theta'\sigma/\sqrt{1-\sigma^2}$. The former equals $R d\phi \rho d\theta'(1-\sigma\cos\theta)^2/\sqrt{1-\sigma^2}$ from (3), so the area ratio is $(1-\sigma\cos\theta)^2$. Since circulation (the line integral of velocity around a loop) is unchanged in the mapping, the vorticity (infinitesimal circulation/infinitesimal area) is mapped with that area ratio factor. Using (2) and (4) the vorticity on the rectangle is

$$2\omega_\perp \cos\theta \cos\phi\, (1-\sigma\cos\theta)^2 = 2\omega_\perp \cos\phi\, \frac{\sigma+\cos\theta'}{(1+\sigma\cos\theta')^3} \qquad (5)$$

*Vortex pair lattice flow*

The vorticity distribution (5) on the periodic rectangle is to be built up as the weight factor in a distribution of shifted point vortex lattices, identical rows each having alternating ±δ-functions. The key point of this convolution, or 'Green function', approach is that an individual lattice of point vortices has a known associated flow field (fig 7). This flow involves a special function, the Elliptic function sn. However it is only invoked temporarily since when convolved with the vorticity distribution, the outcome is expressed in terms of elementary functions, so in the end no reference to the intricacies of Elliptic functions is required.

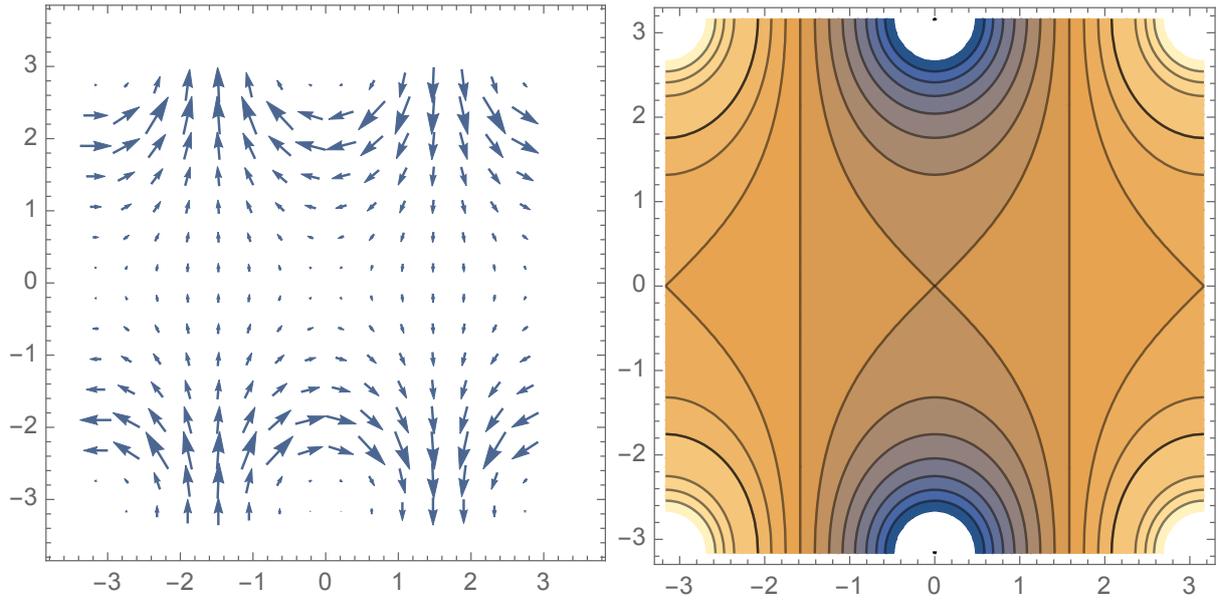

Fig 7. Two equivalent pictures of the flow associated with a lattice of fixed point vortices. Every unit cell contains a single positive vortex (one quarter vortex in each corner) and a single negative vortex (two halves). Left: The flow velocity vector field (Im(sn), Re(sn)) where sn is the Elliptic function in a periodic rectangle (here pictured for the 'modulus' parameter $k$ for which the rectangle is square) that has zero div everywhere, and zero curl except at the two point vortices of opposite signs (±δ-functions of curl). Right: The corresponding streamlines are contours of the real part of the integral of sn, explicitly $\mathrm{Re}\int_0^z \mathrm{sn}(z,k)\,dz$=Re($k^{-1}$ log(dn(z,k)−$k$ cn(z,k))).

For the present periodic rectangle (deriving from the torus) with coordinates ($\phi,\theta'$), one takes $z=(2K\phi+iK'\theta')/\pi$, with $K(k)$ and $K'(k)$ being the standard complete Elliptic integrals with $k$ such that $K'/2K = \sigma/\sqrt{1-\sigma^2}$, the rectangle aspect ratio where $\sigma=\rho/R$. The appropriate streamfunction $s_{\delta\delta}$ is then simply proportional to Re($\int$sn) with a prefactor such that $\nabla^2 s_{\delta\delta} = \pm\delta/2$ at the vortex points. The 2 is needed because there is a pair of opposite vortices per rectangle and each will contribute equally to the convolution because of the antisymmetry of the vorticity function. Since the residue of the poles of sn are ±1/$k$, the circulation around each is ±2π/$k$. Thus the desired streamfunction with the point vortices solving the above Laplacian equation is $s_{\delta\delta} = (k/4\pi)$ Re($\int$sn), written out more fully in (6), (7) and (8) below. This is both a function in the complex $z$ plane (periodic with dimensions 4$K$×2$K'$) and, via the link above of $z$ to ($\phi,\theta'$), in the periodic rectangle (dimensions $2\pi R \times 2\pi R\,\sigma/\sqrt{1-\sigma^2}$). These rectangles differ only in scale (not aspect ratio). The corresponding velocity fields are related by the reciprocal of the scaling factor, and thus they have identical circulation values around corresponding arbitrary test circuits, equal to half the count of vortices enclosed.

*Convolution*

The convolution (between the vorticity distribution and the vortex lattice flow on the rectangle) will eliminate the Elliptic function. This comes about using an identity that can be proved by using the Fourier transform product rule for convolutions, and the known Fourier series [Lawden 1989] of the function sn. In this series only the first term gives a non-zero contribution: $\text{sn}(u) = \frac{2\pi\sqrt{q}}{kK(1+q)}\sin(\pi u/2K) + \cdots$, where $q = \exp[-2\pi\sigma/\sqrt{1-\sigma^2}]$, and as always $\sigma = \rho/R$. The identity, with $z = (2K\varphi + iK'\vartheta')/\pi$ where $-\pi < \vartheta' < \pi$, and the curly forms $\varphi, \vartheta'$ of $\phi, \theta'$ used as integration variables henceforth, then reads:

$$\int_{-\pi}^{\pi} s_{\delta\delta}(\varphi, \vartheta') \cos(\phi - \varphi) \, d\varphi \tag{6}$$

$$= \left(\frac{k}{4\pi}\right) \int_{-\pi}^{\pi} \left(\int_0^z \text{sn}(z, k) dz\right) \cos(\phi - \varphi) \, d\varphi \tag{7}$$

$$\left\{ = \left(\frac{k}{4\pi}\right) \int_{-\pi}^{\pi} \left(\frac{1}{k}\right) \log[\text{dn}(z,k) - k\,\text{cn}(z,k)] \cos[\phi - \varphi] \, d\varphi \right\} \tag{8}$$

$$= \frac{\sqrt{q}}{1-q} \cos\phi \cosh\left[\vartheta' \frac{\sigma}{\sqrt{1-\sigma^2}}\right] \tag{9}$$

The line in braces (8) is not essential because sn can be replaced by its first Fourier term, as mentioned, leading directly to the next line, but it is included for explicitness. A shift now needs to be taken into account. The point vortices of $s_{\delta\delta}$ (fig 6), are offset vertically, none is at the origin. But for the purposes of the convolution a vortex needs to be located at the origin. The required shift is $\vartheta' \to \vartheta' - \pi$ so that $s_{\delta\delta}(\varphi, \vartheta')$ becomes $s_{\delta\delta}(\varphi, \vartheta' - \pi)$. It would be incorrect, however, simply to make that substitution in (9), because that has assumed that $-\pi < \vartheta' < \pi$. Outside this range the cosh is truncated and repeated periodically. However the $\pi$ shift can be accommodated by transfer to the convolving partner (5).

The 2D convolution of $s_{\delta\delta} = (k/4\pi)\,\text{Re}(\int\text{sn})$ with the vorticity distribution (5) supplying the stream function $s(\phi, \theta')$ on the periodic rectangle for the diametral turning can finally be written down using (9) and (5). There is a Jacobian factor $R^2\sigma/\sqrt{1-\sigma^2}$ converting between this $2\pi R \times 2\pi R\sigma/\sqrt{1-\sigma^2}$ rectangle and the $2\pi \times 2\pi$ range of $(\phi, \theta')$.

$s(\phi, \theta')$

$$= \left(\frac{k}{4\pi}\right) \int_{-\pi}^{\pi} \int_{-\pi}^{\pi} s_{\delta\delta}(\varphi, \theta')$$

$$\times \cos[\phi - \varphi] \, 2\omega_\perp \cos\varphi \frac{\sigma + \cos(\theta' - \vartheta' - \pi)}{(1 + \sigma\cos(\theta' - \vartheta' - \pi))^3} R^2 \frac{\sigma}{\sqrt{1-\sigma^2}} d\varphi d\vartheta' \tag{10}$$

$$= 2\omega_\perp R^2 \frac{\sigma}{\sqrt{1-\sigma^2}} \frac{\sqrt{q}}{1-q} \cos\phi \int_{-\pi}^{\pi} \cosh\left[\vartheta' \frac{\sigma}{\sqrt{1-\sigma^2}}\right] \frac{\sigma - \cos(\theta' - \vartheta')}{(1 - \sigma\cos(\theta' - \theta'))^3} d\vartheta' \tag{11}$$

Then the stream function on the torus is

$$s(\phi, \theta'(\theta)) = s\left(\phi, 2\tan^{-1}\left(\frac{\sqrt{1+\sigma}}{\sqrt{1-\sigma}}\tan\frac{\theta}{2}\right)\right) \qquad (12)$$

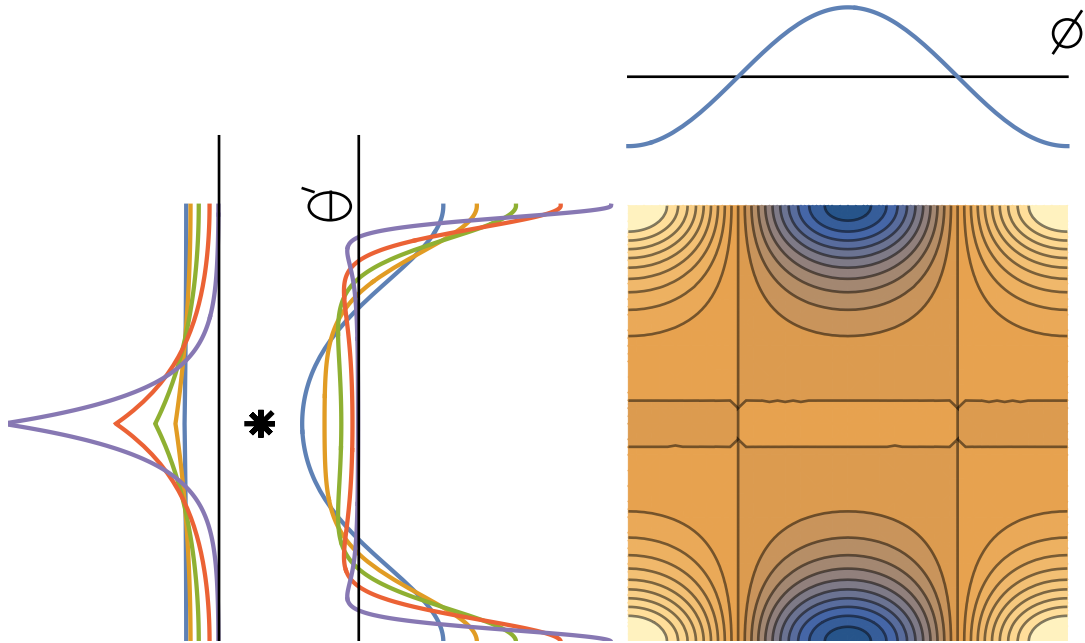

Fig 8. Construction of the flow streamlines on the (conformally mapped) torus turning about a diametral axis; the graphical interpretation of the algebraic formulas (10) and (11). A particular case $\sigma=\rho/R=0.7$ is chosen here for the contour map (of the flow streamfunction). The streamfunction is the product of a simple cosine horizontally ($\phi$ axis), and the convolution of the two red functions vertically ($\theta'$ axis). The other colours correspond to different values $\sigma$ = 0.1, 0.3, 0.5, and 0.9. All of the vorticity graphs (the right hand set) have two zeros, but when convolved with the chopped cosh graphs (left hand set) only the sharper pointed ones (0.9 and 0.7) retain two zeros, yielding the twin horizontal streamlines shown for the 0.7 case. The less sharp ones smear away the zeros and would yield a product contour map with two diagonal crosses (separatrices) like those in fig 7 (and, on the torus in fig 3).

*Remarks*

(i) For the extremes of thin ($\sigma=0$) and fat ($\sigma=1$) tori, the convolution integral (12) admit analytic evaluation. These evaluations will not be carried out, but the appropriate approximations are as follows. The thin limit comes from straightforward Taylor expansion in $\sigma$ in the integrand of (12). The fat limit comes from expanding the two functions being convolved in (12) about their peaks, namely the gradient discontinuity for the cosh, and the $\pi$ argument of the cos. Only the constant, and first non-constant, terms are required, and lead to integrals consisting of exponential decays times simple rational functions, and therefore resulting in exponential integral special functions like Ei. The qualitative behaviour in this fat limit is interesting. The peak widths of the two convolved functions are both of order $\sqrt{\varepsilon}$ where $\varepsilon=1-\sigma$, so the convolution is the same order $\sqrt{\varepsilon}$ in width. But these are small widths about $\theta'=\pi$ on the rectangle, not on the torus. To convert to the torus one needs the mapping (4) and this undoes the smallness since $\sqrt{1-\sigma}/\sqrt{1+\sigma} \sim \frac{\sqrt{\varepsilon}}{2}$. (in the relativistic aberration context this is called the headlight effect). The result is that the features of the flow, importantly the twin rings in fig 1 and fig 4, are not in extreme locations on the torus.

(ii) The surprise that any surface shape possesses an intrinsic wall to flow induced by turning is perhaps emphasised by the fact that both more symmetric and less symmetric surfaces lack it. A sphere lacks it, any chosen initial point can be driven to any other chosen position by a suitable sequence of rotations. Even a spheroid with its axial symmetry like the torus does not exhibit an intrinsic wall because it lacks the circles along which the surface normal is parallel to the axis. A torus is geometrically generated by swinging a circle about an axis in its plane, but the wall phenomenon would still be present with a deformed circle instead (though the conformal mapping to a rectangle will be less straightforward [Guenther 2020]). Also the surface of a thick bowl might posess an intrinsic wall. An ellipsoid with distinct axis lengths would not because there would be no streamline common to all three rotation axes. The same would seem to apply to even less simple non symmetric surfaces.

(iii) An aspect of the reconfiguration of the liquid that has not been mentioned is angle holonomy. It is quite separate from the wall phenomenon except insofar as it also derives from the geometric nature (timing irrelevance) of the reconfiguration. Angle holonomy in classical mechanics is restricted to an integrable system that is subjected to an adiabatic excursion of its Hamiltonian [Hannay 1985]. (Actually the arena of integrable systems is also tori, but in phase space, and not directly connected with the present real space torus). A simple example is a bead gliding frictionlessly around a rigid wire ring or hoop. The hoop need not be planar, though the easiest example is a circular one. If the hoop is manipulated so that it ends up where it started, the bead will have acquired an extra shift beyond where it would have travelled to had the hoop not moved at all. If the manipulation is adiabatic (slow compared to the bead's sliding circulation) the extra shift (in arc length) equals the solid angle swept out by the area vector, times twice its magnitude, and divided by the hoop's perimeter. (The area vector is half the integral of position vector cross unit tangent vector darclength). This same shift is experienced by a perfect liquid in a thin rigid hollow hoop. Now however there is no adiabaticity condition, the liquid can start stationary – it is a 1D flow version of the present 2D flow problem, and one should expect an analogous shift holonomy to apply to the torus. Treating the torus manipulation as infinitesimal alternating diametral and axial rotations, it is indeed clear that only the latter matter, and the area averaged azimuthal liquid coordinate $\phi$ shifts by the solid angle swept out by the torus axis, in accordance with the 1D formula applied to a circular hoop.

(iv) For the more general problem of bulk 3D flow instead of the surface flow considered here, there is some connection with previous study of the external flow problem – 'swimming' in a perfect liquid [Saffman 1967] [Hannay 2012]. The 'swimmer' could be taken as rigid object (here the empty container) immersed in the liquid filling all the outside space.

The object is manipulated, perhaps by moving a mass around inside. Despite much similarity with the internal flow problem, an obvious difference is that zero container angular velocity vector does not mean that the liquid is stationary in the container frame of reference frame, the container velocity vector matters too.

(v) Returning to a contained perfect liquid in bulk 3D, there are similarities and differences with the surface flow problem. Only rotations matter, and each component of container angular velocity has an associated basis flow, the three of them superposing linearly. Each flow has uniform relative vorticity: minus twice the relevant resolved component of angular velocity. Each is divless but in 3D there is no streamfunction to simplify analysis. A virtual wall would be a surface to which all three basis flow velocity fields were tangent.